\documentclass[aps,preprint,prx,superscriptaddress,longbibliography,showkeys]{revtex4-1}
\usepackage{graphicx}
\usepackage{hyperref}
\usepackage{amsmath}
\usepackage{amssymb}
\usepackage[usenames,dvipsnames]{color}
\usepackage{ragged2e}

\begin{document}

\title{Synthesis and Local Probe Gating of a Monolayer Metal-Organic Framework}

\date{\today}

\author{Linghao Yan}
\email{Email: linghao.yan@aalto.fi}
\affiliation{Department of Applied Physics, Aalto University, FI-00076 Aalto, Finland}

\author{Orlando J. Silveira}
\affiliation{Department of Applied Physics, Aalto University, FI-00076 Aalto, Finland}

\author{Benjamin Alldritt}
\affiliation{Department of Applied Physics, Aalto University, FI-00076 Aalto, Finland}

\author{Ond\v{r}ej Krej\v{c}\'i}
\affiliation{Department of Applied Physics, Aalto University, FI-00076 Aalto, Finland}

\author{Adam S. Foster}
\affiliation{Department of Applied Physics, Aalto University, FI-00076 Aalto, Finland}
\affiliation{Nano Life Science Institute (WPI-NanoLSI), Kanazawa University, Kakuma-machi, Kanazawa 920-1192, Japan}

\author{Peter Liljeroth}
\email{Email: peter.liljeroth@aalto.fi}
\affiliation{Department of Applied Physics, Aalto University, FI-00076 Aalto, Finland}


\begin{abstract}
Achieving large-area uniform two-dimensional (2D) metal-organic frameworks (MOFs) and controlling their electronic properties on inert surfaces is a big step towards future applications in electronic
devices. Here we successfully fabricated a 2D monolayer Cu-dicyanoanthracene (DCA) MOF with long-range order on an epitaxial graphene surface. Its structural and electronic properties are studied by low-temperature scanning tunneling microscopy (STM) and spectroscopy (STS) complemented by density-functional theory (DFT) calculations. We demonstrate access to multiple molecular charge states in the 2D MOF using tip-induced local electric fields. We expect that a similar strategy could be applied to fabricate and characterize 2D MOFs with exotic, engineered electronic states.
\end{abstract}

\maketitle

\section{Introduction}
Metal-organic frameworks (MOFs) are an important class of materials that have been intensively studied in the last two decades. Despite the vast number of reports on three-dimensional, bulk MOFs, synthesis and characterization of two-dimensional (2D), single layer MOFs are much more limited \cite{Kambe2013-ConjugatedNanosheet,Gao2019SynthesisGap,Zhang2020OnsurfaceBands}. Intrinsic 2D MOFs are expected to attract increasing attention since they are anticipated to possess exotic electronic properties, such as high electrical conductivity \cite{Sheberla2014HighAnalogue,Kambe2014RedoxInsulator,Huang2015ABehaviour,Chen2018HighlyBenzenehexathiol,Xie2020ElectricallyFrameworks}, superconductivity \cite{Zhang2017TheoreticalFramework,Huang2018SuperconductivityStructure}, topologically non-trivial band structure \cite{Wang2013OrganicLattices,Jiang2017,CrastodeLima2019LayertronicFrameworks,Zhang2019PredictionFrameworks,Jiang2020TopologicalFrameworks,Silveira2020Activationsub4/sub,Gao2020QuantumLattice}, half-metallic ferromagnetism \cite{Zhao2013Half-metallicityMonolayer,Zhang2015RobustCu-TPyB,Jin2018Large-gapFrameworks,Zhang2019Two-dimensionalLattice} and quantum spin liquids \cite{Yamada2017DesigningFrameworks}. To isolate their intrinsic electronic properties from the substrate, synthesis of 2D MOFs on inert surfaces, such as graphene, other van der Waals layered materials, and bulk insulators, is highly desired. Furthermore, understanding the performance of 2D MOFs in a gated device at the atomic scale would be essential for future applications.

On-surface 2D porous metal-organic networks, which represent a 2D analogue of 3D MOFs, have been fabricated following the concepts of supramolecular coordination chemistry \cite{Lin2009,Barth2009FreshChemistry}. By tuning the selection of metal atoms and organic linkers, different kinds of lattice structures of 2D MOFs can be fabricated \cite{Dong2016Self-assemblySurfaces}. Since the symmetry of the band structure is controlled by the lattice structure, it is expected that specific electronic properties can be realized in certain MOF geometries. However, there are very few reports on 2D MOFs on inert surfaces, where the MOF would retain its intrinsic electronic properties. More crucially, existing work only demonstrates order on a local scale with very small domain sizes of the MOF \cite{Kambe2013-ConjugatedNanosheet,Urgel2015ControllingMonolayer,Schuller2016,Kumar2018Two-DimensionalFrameworks,Zhao2019OnSurfaceTemplates,Li2019}. The characterization of single layer MOFs in a gated device environment has not been realized yet, since it is even more challenging to fabricate 2D MOFs on a three-terminal device. However, this could be alternatively achieved by gating the 2D MOFs using local electric fields induced by the tip of a scanning probe microscope. 

In this work, we successfully fabricated 2D monolayer Cu-dicyanoanthracene (DCA) MOF on an epitaxial graphene surface under ultra-high vacuum (UHV) conditions by precisely controlling the growth parameters. Its structural and electronic properties are studied by low-temperature scanning tunneling microscopy (STM) and spectroscopy (STS) complemented by density-functional theory (DFT) calculations. The ordered DCA$_3$Cu$_2$ network shows a structure combining a honeycomb lattice of Cu atoms with a kagome lattice of DCA molecules and is predicted to be a 2D topological insulator \cite{Zhang2016IntrinsicLattices}. Notably, we demonstrate very long-range order of the DCA-Cu MOF and show that it can grow across the step-edges of the underlying substrate. This facilitates the synthesis of uniform, single-crystalline monolayer MOFs. Combing the STM/STS data with DFT results, we confirm that a kagome band structure is formed in the 2D MOF near the Fermi level. Interestingly, we found that multiple molecular charge states can be generated and modified through the tip-induced local electric fields.

\section{Results and Discussion}
We deposit DCA molecules and Cu atoms sequentially onto the graphene (G/Ir(111)) substrate held at room temperature (details given in the Experimental Section). By tuning the DCA:Cu ratio, both the DCA$_3$Cu single complex (Figures~S1  and S11) and the DCA$_3$Cu$_2$ honeycomb network can be fabricated. The initial sample quality can be improved by annealing the sample at 50 $^\circ$C. This helps to grow larger DCA$_3$Cu$_2$ networks up to a full monolayer, as shown in Figures~1 and S2a. The proper annealing temperature is vital for the formation of a uniform monolayer 2D MOF, since the coordination bonds are relatively weak and reversible \cite{Shi2011ThermodynamicsScale,Cai2017CompetitionCoverage}. The networks are structurally robust, which is evidenced by the fact that they grow seamlessly across step edges of the underlying Ir(111) substrate (see Figure 1b). Annealing at a higher temperature ($>70^\circ$C) yields a new phase of close-packed DCA molecules and large Cu islands (see Figure~S2b). 

\begin{figure}
    \centering
    \includegraphics[width=.9\textwidth]{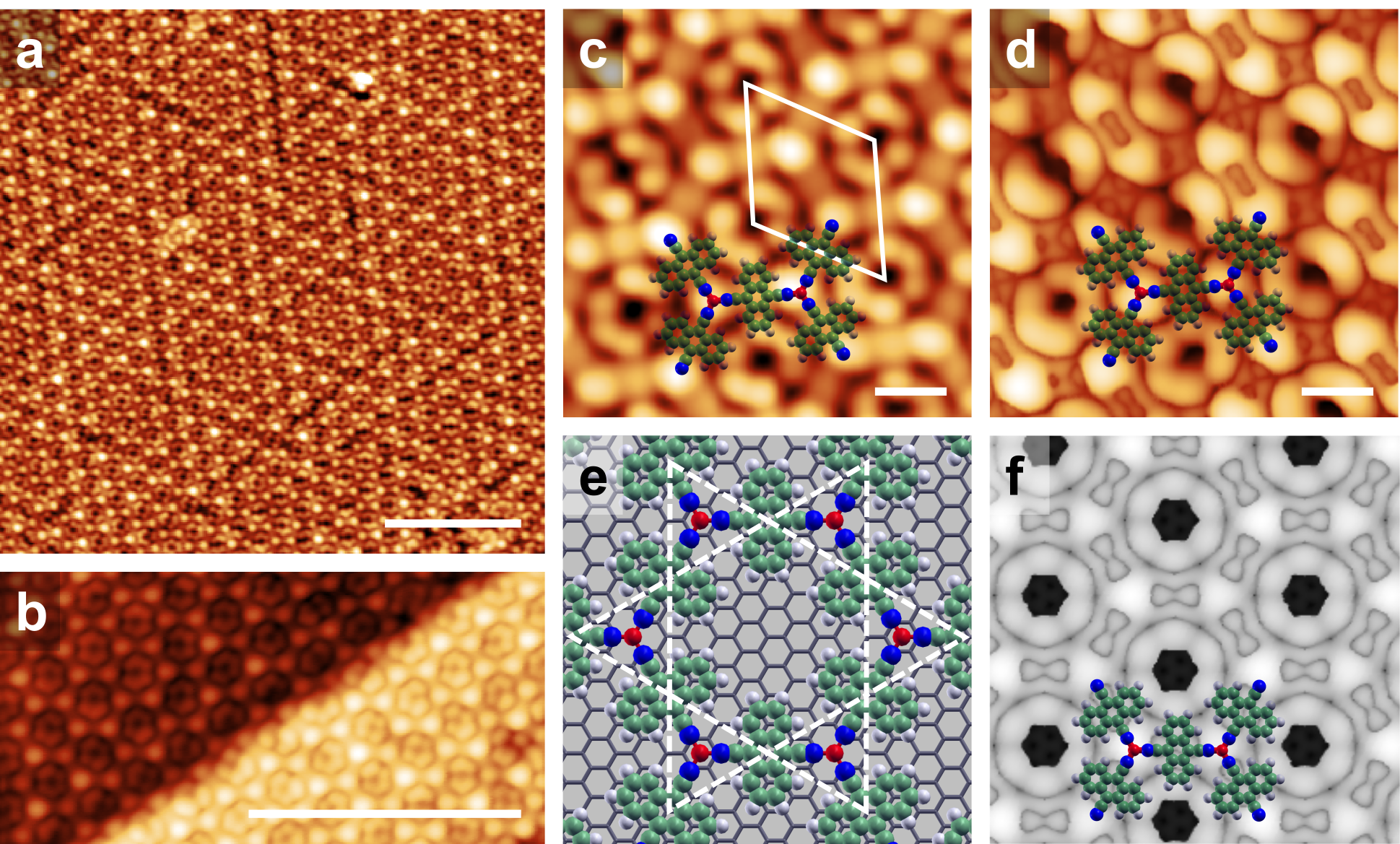}
    \caption{(a) An STM overview image of a DCA$_3$Cu$_2$ MOF on G/Ir(111) surface. (b) An STM image of a DCA$_3$Cu$_2$ MOF seamlessly across step edges of the underlying G/Ir(111) substrate. (c) STM topography image of DCA$_3$Cu$_2$ MOF. The white parallelogram indicates the unit cell. (d) STM image of DCA$_3$Cu$_2$ MOF corresponding to the low energy electronic band of the network. (e) DFT-simulated structure of DCA$_3$Cu$_2$ MOF on graphene. The white dash lines indicates the kagome array of DCA molecules. (f) DFT-simulated STM image of DCA$_3$Cu$_2$ MOF on graphene obtained using the Tersoff-Hamann approximation \cite{PhysRevB.31.805}. Imaging parameters: (a) 1 V and 10 pA, (b) 0.4 V and 10 pA, (c) 1 V and 10 pA, (d) 10 mV and 10 pA. Scale bars: (a) and (b) 10 nm, (c) and (d) 1 nm.}
    \label{fig:1}
\end{figure}

Figure 1c shows a high-resolution image of the backbone of DCA$_3$Cu$_2$ network, with the unit cell shown as a white parallelogram with a lattice constant of $a=1.98\pm0.03$ nm, which is consistent with the DFT value of 1.98 nm and in the range of previous reports of the network on a Cu(111) surface \cite{Pawin2008AExcess,Zhang2014ProbingNetwork}. While the Cu atoms arrange in a honeycomb lattice in the network, the DCA molecules form a kagome pattern, as shown in the model of Figure~1e. Figure 1d shows the same area at a bias voltage corresponding to the low energy electronic band of the Cu-DCA network (see below for more detailed spectroscopy of the Cu-DCA network electronic structure). The image is acquired with a molecule-modified tip apex, which enhances the spatial resolution of the local density of states (LDOS) \cite{Repp2005MoleculesOrbitals}. The experimental image is nicely reproduced by the STM image simulations based on density-functional theory (DFT) calculations of the Cu-DCA structure shown in Figure~1f (see below for details). The modest contrast difference between the DCA molecules in Figure~1d is caused by the Moir\'e pattern of graphene on Ir(111) (cf. Supporting Information, Figure~S3).

Figure 2 shows the d$I$/d$V$ spectra recorded on different high symmetry sites of the network. All the spectra in Figure~2b exhibit a broad peak in the energy range between 0 V to 0.5 V. The contrast in constant height d$I$/d$V$ maps in Figure~2c is not strongly bias dependent in the range from 0.1 V to 0.4 V and reproduced well by the DFT simulations (Figure~S9). We attribute these features to the band structure formed in the 2D network, which has been well studied in similar DCA$_3$Co$_2$ network \cite{Kumar2018Two-DimensionalFrameworks} and will be explored in more detail in Figure~3. Interestingly, the STS measured on the center of the DCA molecule (black curve in Figure~2b) shows two sharp dips around -0.6 V and -1.2 V; the STS on top of the Cu atom (green curve in Figure~2b) shows  a small dip around -0.6 V and a sharp peak around -1.2 V; the end of the long axis of the DCA molecule (blue curve in Figure~2b) shows a sharp peak around -0.6 V and a tiny dip around -1.2 V. The peaks/dips at these two bias values are attributed to the typical charging features, where the charge state of the molecule under the tip changes due to the tip-induced local electric field \cite{Wu2004ControlTransport,Nazin2005VibrationalCrystal,Pradhan2005AtomicCharging,Nazin2005ChargingCrystal,Fernandez-Torrente2012GatingFields,Schulz2013TemplatedNitride,Schulz2015Many-bodyMicroscopy,Liu2015InterplayInsulator,Wickenburg2016,Kocic2019ImplementingMonolayers,Kumar2019ElectricNanoarray,Pham2019SelectiveDoping,Portner2020}. We will discuss the details of these charging features in Figure~4. The spectra of the MOF on the step edge of the underlying Ir(111) substrate (Figure~S12) is consistent with the one on the flat area (Figure~2), indicating that the electronic properties of the MOF are effectively decoupled from the metal substrate by the graphene layer. Besides, all the spectra show a small and sharp peak at or very close to the Fermi level (0-10 mV), which we attribute as a charging peak as well (cf.~Supporting Information, Figure~S4).

\begin{figure}
    \centering
    \includegraphics[width=.9\textwidth]{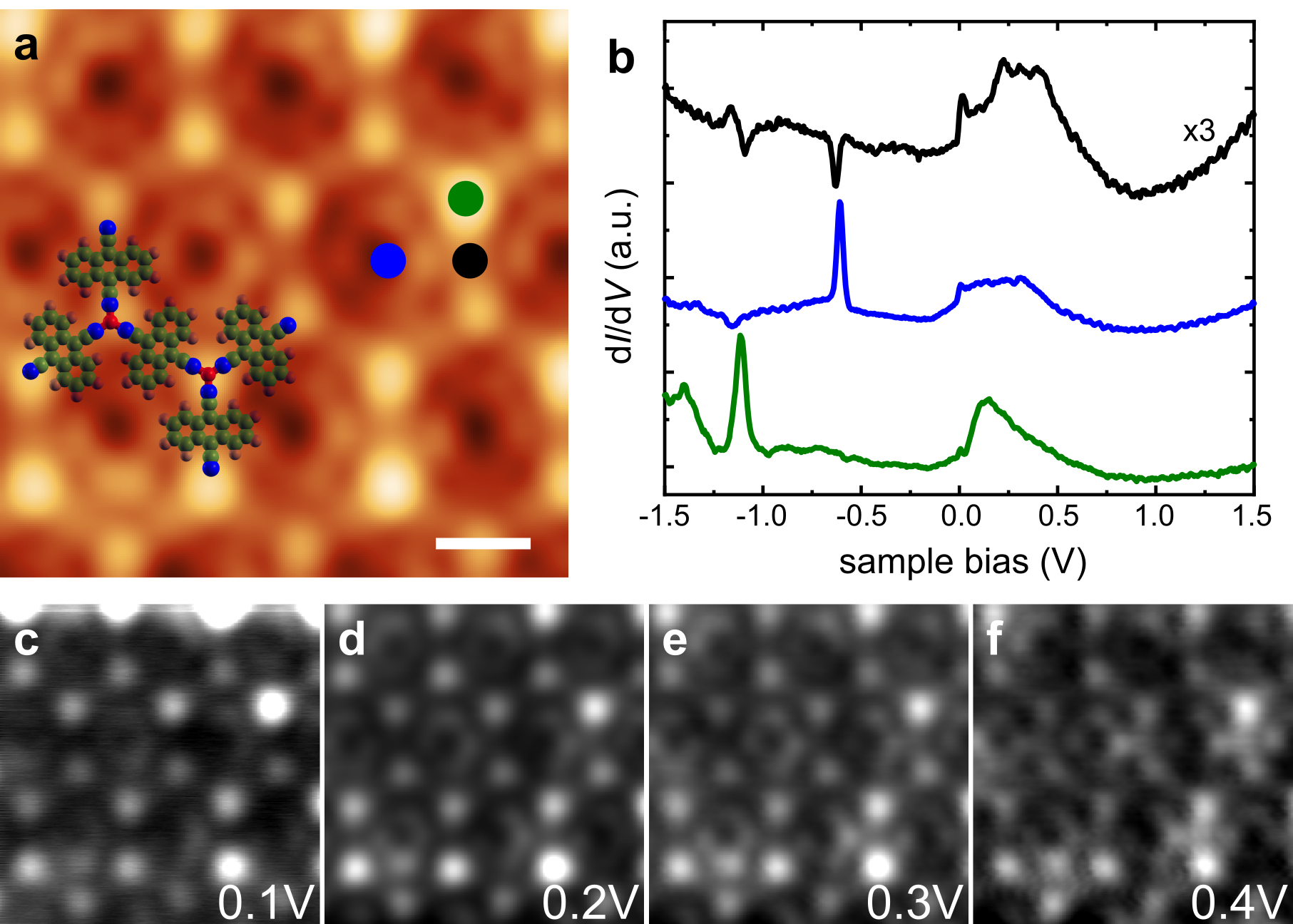}
    \caption{(a,b) STS recorded (b) on DCA$_3$Cu$_2$ network at the positions shown in (a), the spectra were vertically offset for clarity, the spectrum in black has been magnified by a factor of 3 to highlight the weak features. Imaging parameters: 1.5 V and 10 pA. Scale bar: 1 nm. (c-f) Experimentally recorded constant-height d$I$/d$V$ maps at the energies indicated in the panels in the same area of (a).}
    \label{fig:2}
\end{figure}

The DFT simulated band structure of the gas-phase DCA$_3$Cu$_2$ network (Figure~3a) shows a kagome band structure around Fermi level which consists of a Dirac band with an additional flat band pinned to the top of the Dirac band \cite{Leykam2018a,Yan2019EngineeredNanoribbons,Jing2019Two-DimensionalSemiconductors}. The band structure of the DCA$_3$Cu$_2$ network on graphene (the MOF states represented in purple in Figure~3b) shows very similar features with an additional avoided crossing between the MOF kagome and the graphene Dirac bands due to weak hybridization between them. While the Dirac points of the isolated DCA$_3$Cu$_2$ network and graphene are both located at the Fermi level in the gas phase, the Dirac point in the DCA$_3$Cu$_2$ kagome band on graphene is 0.3 eV above the Dirac point of the graphene substrate, indicating a charge transfer between the DCA$_3$Cu$_2$ network and the graphene substrate (cf. Figure~S7). This is consistent with the experimental finding that the energy levels of the DCA$_3$Cu$_2$ kagome band are mostly above the Fermi level. Note that in a clean G/Ir(111) sample, the Dirac point of graphene is 100 meV above the Fermi level as well \cite{Pletikosic2009DiracIr111}. However, the experimental d$I$/d$V$ spectra did not resolve the bandgap of the MOF at the Dirac point. This is mainly due to a certain lifetime broadening and the satellite vibrionic peaks coming from the intermolecular electronic coupling which broadens the spectra.

\begin{figure}
    \centering
    \includegraphics[width=0.9\textwidth]{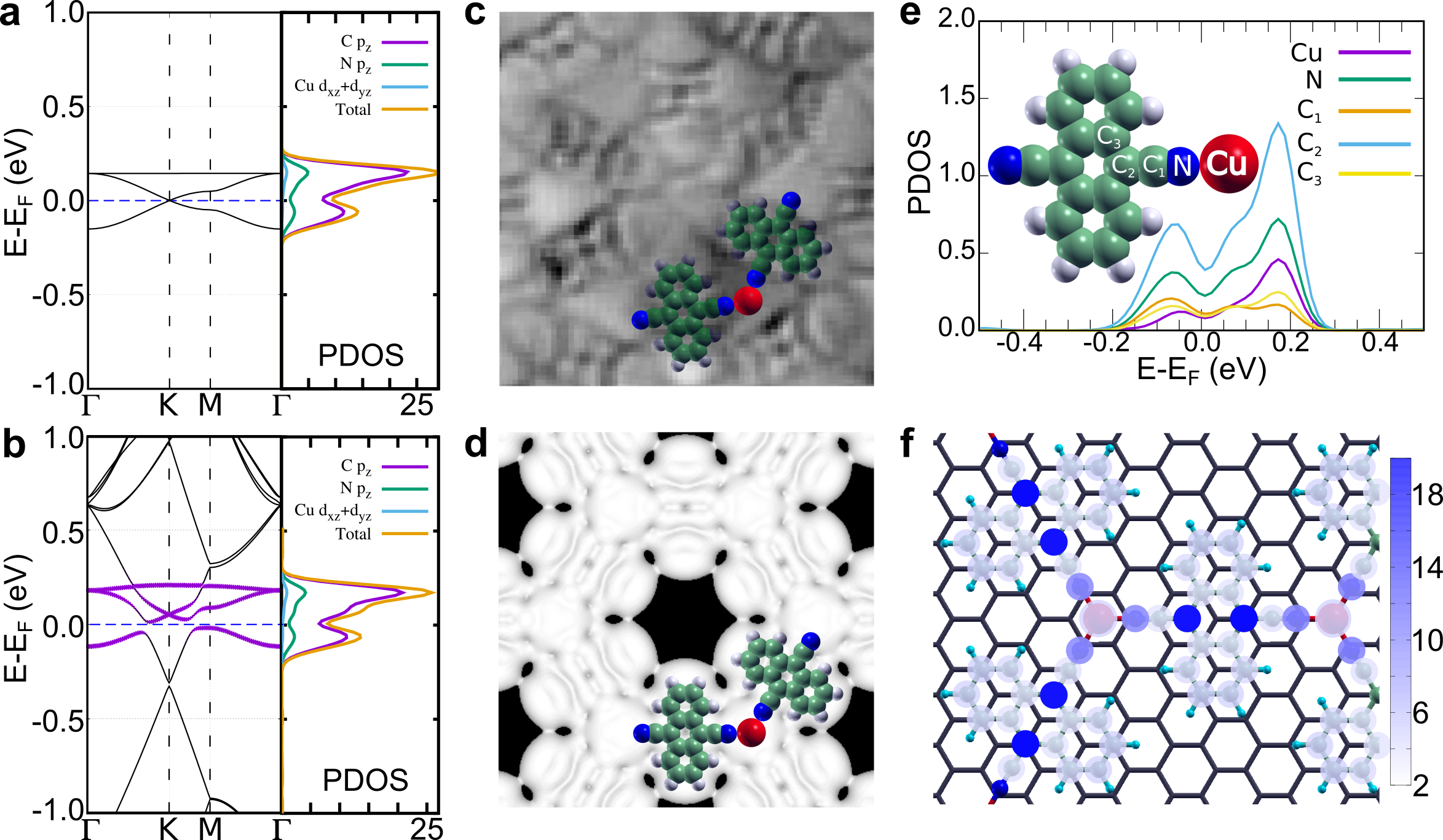}
    \caption{(a) Calculated band structure and PDOS of gas-phase DCA$_3$Cu$_2$ MOF. (b) Calculated band structure and PDOS of DCA$_3$Cu$_2$ MOF on graphene. The purple color in the band structure indicates the contribution from the MOF's electrons. (c,d) Examples of the (c) experimental and (d) simulated LDOS maps. The bias voltage of the map in (c) is 0.2 V and the energy in (d) is 0.1 eV, obtained considering a mixed $sp$-wave tip (5\% of $s$ and 95\% of $p$). (e) PDOS showing the contribution of selected atoms of the MOF to the kagome band. The full energy range of this plot is used for the integrated density contribution shown in (f), where the color intensity of the blue disks overlaid above the structure depicts the atomic contribution to the integrated PDOS. Both (e) and (f) show that the kagome states are located mainly on the central carbon and nitrogen atom of DCA molecules, with only minor contributions from Cu.}
    \label{fig:3}
\end{figure}

The projected density of states (PDOS) in Figure~3 show that the kagome band originates mainly from the DCA molecule ($p_{z}$ orbitals of C and N atoms), with very minor contributions from the Cu $d$ orbitals ($d_{xz}$+$d_{yz}$), indicating that the kagome geometry of the DCA molecular array and the $\pi$-$d$ extended conjugation in the metal-organic framework is realized in this band structure \cite{Sheberla2014HighAnalogue,Huang2015ABehaviour,Dou2017SignatureCu,Day2019SingleProperties,Xie2020ElectricallyFrameworks}. The DFT simulated LDOS maps show uniform features at different energies within the kagome band (Figures~S5a-d). To probe the electronic structure of the DCA$_3$Cu$_2$ network in more detail, we used a molecule-modified $p$-wave tip to get the high-resolution LDOS maps \cite{Gross2011High-resolutionTip} shown in Figures~S5e-h. These show homogeneous appearances which are similar at different biases as in the simulated results.
The representative experimental and DFT simulated LDOS maps are shown in Figures~3c and 3d, respectively. The experimental maps match the representative DFT simulated data very well. The minor contrast difference comes from the moir\'e pattern and slight inhomogeneities of the graphene substrate, which is also reflected in the charging rings shown in Figure~4.

Charging behavior has been studied in detail in the case of single molecules or self-assembled molecular monolayers on coinage metal surfaces such as Au(111) \cite{Fernandez-Torrente2012GatingFields} and Ag(111) \cite{Kocic2019ImplementingMonolayers,Kumar2019ElectricNanoarray} and more frequently found when the molecule is decoupled from the substrate by an ultra-thin film such as Al$_2$O$_3$ \cite{Wu2004ControlTransport,Nazin2005VibrationalCrystal,Pradhan2005AtomicCharging,Nazin2005ChargingCrystal}, graphene \cite{Wickenburg2016, Pham2019SelectiveDoping} and hexagonal boron nitride \cite{Schulz2013TemplatedNitride, Schulz2015Many-bodyMicroscopy,Liu2015InterplayInsulator,Portner2020}. It can be explained by considering the tip-molecule-substrate system as a double-barrier tunnel junction (DBTJ) as illustrated in Figures~4a and 4b. When a bias voltage is applied across the DBTJ, there is a potential distribution with a drop at both the tip-molecule and the molecule-substrate junctions. Consequently, applying a bias voltage $V_\mathrm{b}$ causes the molecular levels to shift in energy by $\alpha V_\mathrm{b}$, where $\alpha$ is the fraction of the potential drop between the molecule and the substrate with respect to the overall bias. If there are molecular energy levels close to the Fermi level, these can shift across it at some value of the bias voltage and the charge state of the molecule changes by $\pm e$. In our particular case, the charging occurs once the band bottom of the kagome band (KB) shifts down below the Fermi level: $(1-\alpha) V_\mathrm{b}=E_\mathrm{F}-E_\mathrm{KB}$. The charging features can be distinguished from the usual molecular resonances by checking how the charging peak/dip shifts as a function of the tip-molecule distance. Bringing the tip closer to the molecule reduces $\alpha$ and, consequently, the charging features shift towards the Fermi level monotonously with decreasing tip-molecule distance (shown in Figure~4c),  which is consistent with previous studies \cite{Wu2004ControlTransport,Pham2019SelectiveDoping,Kumar2019ElectricNanoarray}. The charging features shift by about 8.8 mV/\AA~upon reducing the tip-sample distance approach. On the other hand, the kagome band position at the positive bias barely changes with different tip-molecule distances, which can also be seen from Figure~4c. 

\begin{figure}
    \centering
    \includegraphics[width=.9\textwidth]{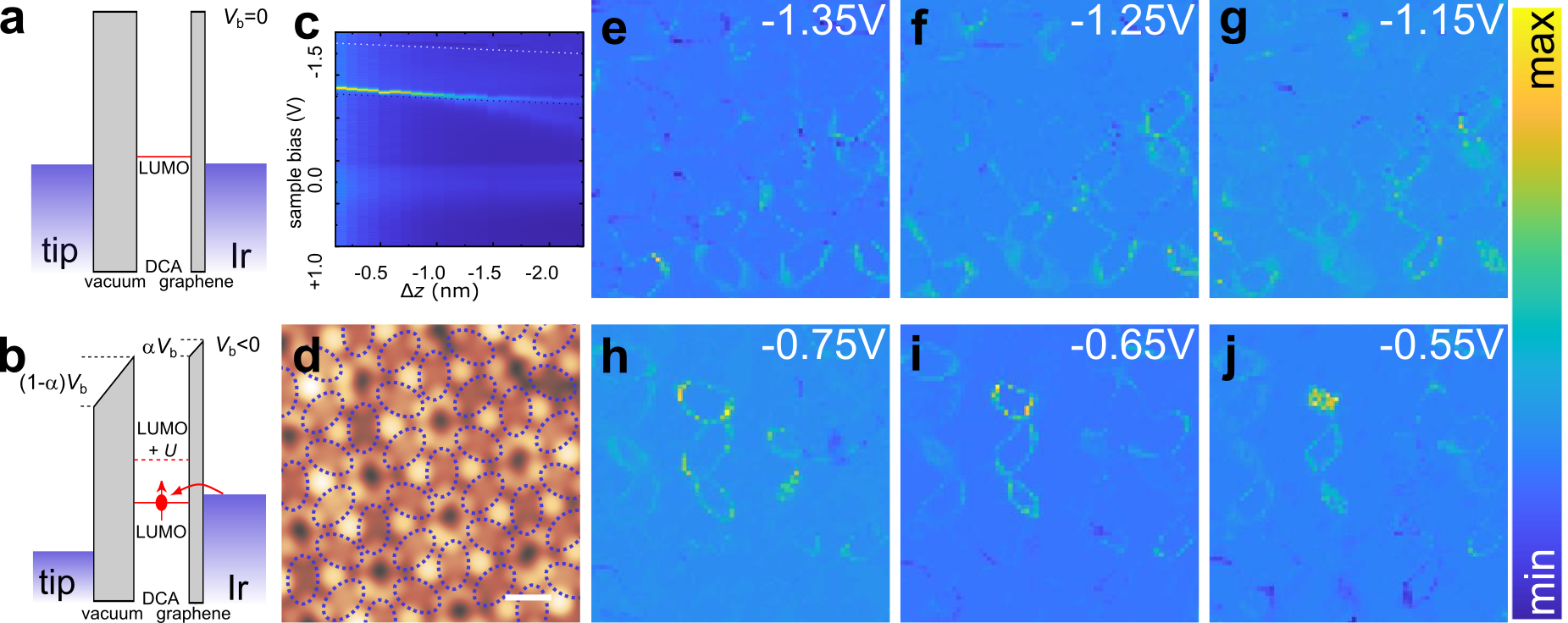}
    \caption{(a,b) Schematics of charging process in the DBTJ model, exhibiting the down shift of the molecular orbital resulting in the charging of the DCA molecules. (c) LDOS map showing the shift of the charging features (highlighted with dotted lines) vs.~the tip-sample distance, $\Delta z$ ($\Delta z<0$ represents reducing tip-sample distance). (d) topography and (e-j) LDOS maps at different bias voltages demonstrating the charging rings in the same area of (d). The dotted purple ellipses in (d) show all the possible charging rings. Imaging parameters of (d): 0.4 V and 10 pA. Scale bar of (d): 1 nm.}
    \label{fig:4}
\end{figure}

Another fingerprint of a charging phenomenon is the charging ring feature. As shown in Figures~4d-4j, the elliptical rings observed in the LDOS maps at different bias voltages represent the onset of charging as the tip is moved towards the charging site. Notably, in the previous experiments, both the charging peaks \cite{Nazin2005VibrationalCrystal,Pradhan2005AtomicCharging,Nazin2005ChargingCrystal,Wickenburg2016,Pham2019SelectiveDoping} and dips \cite{Fernandez-Torrente2012GatingFields,Kocic2019ImplementingMonolayers,Kumar2019ElectricNanoarray} have been found at negative bias. The presence of a peak or dip in the d$I$/d$V$ spectrum depends on whether the LDOS at the tip position increases or decreases due to the charging event, which might also depend on the local adsorption registry between the MOF and the underlying graphene substrate. The charging peak/dip rings features around two different bias voltages: one around -1.2 V (Figures~4e-4g) and the other -0.6 V (Figures~4h-4j). The energy difference of the two charging rings reflects the Coulomb charging energy ($\sim$ 0.6 eV). The peak and dip features can either coexist around a specific molecule or one of them can be dominant. Around both biases, the charging ring perimeter decreases as the bias becomes less negative, which is also consistent with the previous works. Figure~4d highlights all the possible charging rings with dotted purple ellipses. However, the charging ring positions and perimeters in the experimental data are not entirely identical (Figures~4e-4j). Each molecule has slightly varying adsorption environment (e.g.~due to the moir\'e pattern on graphene on Ir(111) which is known to give rise to a work function modulation of a couple of hundred meV \cite{Craes2013MappingDots,Jarvinen2014Self-assemblySurface,Altenburg2014LocalIr111,Kumar2017Charge-Transfer-DrivenGraphene}), varying the exact on-set bias of the charging \cite{Fernandez-Torrente2012GatingFields,Liu2015InterplayInsulator,Kumar2019ElectricNanoarray,Portner2020}. The elliptical rings are mostly around the DCA molecules, which is consistent with the DFT results that the band above the Fermi level (which is pulled below $E_\mathrm{F}$ at negative bias) has most of its density on the DCA  molecules. These local charging features demonstrate that electron-electron interactions (characterized by the Coulomb charging energy,  here $\sim$ 0.6 eV) are significant and they can be expected to be of a similar magnitude compared to the overall band width of the Cu-DCA network (here several hundreds of meV.)). Cu-DCA has been predicted to be an intrinsic 2D topological insulator in a non-interacting model \cite{Zhang2016IntrinsicLattices}, but our results indicate the need to go beyond this simple picture and consider electron-electron interactions.

\section{Conclusion}
In summary, we study the structural and electronic properties of monolayer Cu-DCA MOF on a G/Ir(111) substrate under UHV conditions using experimental (STM/STS) and theoretical (DFT) methods. We demonstrate successful synthesis of a large-scale monolayer MOF that can grow across the terrace of the graphene substrate. The 2D Cu-DCA MOF possesses a kagome band structure near the Fermi level. We expect that a similar strategy could be applied to fabricate and characterize 2D MOFs with exotic electronic states on weakly interacting substrates, e.g. 2D MOFs with heavy metal atoms \cite{Shi2009,Lyu2015On-surfaceStructures,Sun2016DehalogenativeNanostructures,Song2017Self-AssemblySurfaces,Yang2018Two-dimensionalAg111,Yan2018StabilizingNetworks,Sun2018DeconstructionStructure,Yan2019On-SurfaceAu111} possessing strong spin-orbit couplings and the possibility of realizing an organic topological insulator \cite{Springer2020TopologicalPolymers}. In addition, multiple molecular charge states are observed and modified by the tip-induced local electric fields. This highlights the role of electron-electron interactions that are likely to be of a similar order of magnitude as the overall band width. Depending on the relative magnitudes of the different energy scales, this can give rise to magnetically ordered or spin liquid ground states or -- when coupled with spin-orbit interactions -- result in a quantum anomalous Hall insulator or more exotic electronic states. \cite{PhysRevLett.100.156401,pesin2010mott,Ren_2016,Rachel_2018}. 

\section*{Methods}

Sample preparation and STM experiments were carried out in an ultrahigh vacuum system with a base pressure of $\sim 10^{-10}$ mbar. The Ir(111) single crystal sample was cleaned by repeated cycles of Ne+ sputtering at 2 kV and annealing in an oxygen environment at 900 $^\circ$C followed by flashing to 1300 $^\circ$C. Graphene was grown by adsorbing ethylene and flashing the sample to 1100 - 1300 $^\circ$C in a TPG (temperature programmed growth) step followed by a CVD step where the Ir(111) substrate at 1100 - 1300 $^\circ$C is exposed to ethylene gas at $5\times10^{-7}$ mbar pressure for 1 minute \cite{Coraux2009GrowthIr111}. This gives approximately a full monolayer coverage of graphene (G/Ir(111)).

The DCA$_3$Cu single complex and DCA$_3$Cu$_2$ network can be fabricated by the sequential deposition of 9,10-dicyanoanthracene (DCA, Sigma Aldrich) molecules and Cu atoms onto the G/Ir(111) substrate held at room temperature. Further annealing the sample at 50 $^\circ$C results in DCA$_3$Cu$_2$ network growth until monolayer coverage is reached. DCA molecules were thermally evaporated from a resistively heated aluminum oxide crucible at 100 $^\circ$C. Subsequently, the samples were inserted into a low-temperature STM (Createc GmbH), and all subsequent experiments were performed at $T = 5$ K. STM images were taken in the constant current mode. d$I$/d$V$ spectra were recorded by standard lock-in detection while sweeping the sample bias in an open feedback loop configuration, with a peak-to-peak bias modulation of 15 - 20 mV at a frequency of 526 Hz. In the local density of states (LDOS) maps, each d$I$/d$V$ spectrum was normalized by $I/V$ spectra to minimize the height difference contribution \cite{Stroscio1986, Klappenberger2011}. The STM images were processed by Gwyddion software \cite{Necas2012Gwyddion:Analysis}.

The DFT calculations were performed with the QUANTUM-ESPRESSO distribution \cite{Giannozzi_2009}. We used the optB86b-vdW functional \cite{PhysRevB.83.195131, PhysRevLett.92.246401} to optimize the structure of the DCA$_3$Cu$_2$ MOF both as an isolated layer and on graphene.
To describe the interaction between electrons and ions we used PAW pseudopotentials \cite{PhysRevB.50.17953}, while the electronic wave functions were expanded considering a plane-wave basis set with kinetic energy cutoffs of 90 Ry. Integrations over the Brillouin zone (BZ) were performed using a uniform grid of $4 \times 4 \times 1$ $k$-points, and a twice denser grid was used to obtain band structures and PDOS. 
The electronic structure for the STM simulations and LDOS maps were calculated via FHI-AIMS package \cite{aims1} from the previously optimized geometry as a single point calculation. For this calculations we made use of the Perdew-Burke-Ernzerhof (PBE) exchange-correlation functional \cite{PhysRevLett.77.3865} with $\Gamma$ $k$-point only. The \textit{s} and \textit{p$_{xy}$} wave STM simulations and LDOS maps were then computed by means of the PP-STM code with fixed tip, where the broadening parameter $\eta$ was set to 0.3 eV \cite{PPSTM}. The fixed tip simulations fingerprint only the electronic structure of the  sample in the following manner: The \textit{s}-tip d$I$/d$V$ signal is given by: $LDOS_{s}(E,\vec{r_0})\propto\left|\psi(E,\vec{r_0})\right|^2$, while the \textit{p$_{xy}$}-tip dI/dV signal can be calculated as: $LDOS_{p_{xy}}(E,\vec{r_0})\propto\left|\frac{\partial \psi}{\partial x} (E,\vec{r_0}) \right|^2 + \left|\frac{\partial \psi}{\partial y} (E,\vec{r_0}) \right|^2 $ \cite{Gross2011High-resolutionTip,PPSTM}. The normalized LDOS maps have been simulated via plotting iso-surface of the corresponding d$I$/d$V$ signal.

\setlength{\parindent}{0 cm} 

\medskip
\textbf{Supporting Information} \par
Supporting Information is available.

\medskip
\textbf{Acknowledgments} \par
This research made use of the Aalto Nanomicroscopy Center (Aalto NMC) facilities. We acknowledge support from the European Research Council (ERC-2017-AdG no.~788185 ``Artificial Designer Materials'') and Academy of Finland (Academy projects no.~311012 and 314882, and Academy professor funding no.~318995 and 320555). Linghao Yan and Ond\v{r}ej Krej\v{c}\'i acknowledge support from European Union's Horizon 2020 research and innovation program (Marie Skłodowska-Curie Actions Individual Fellowship no.~839242 ``EMOF'' and no.~845060 ``QMKPFM''). Computing resources from the Aalto Science-IT project and CSC, Helsinki are gratefully acknowledged. ASF has been supported by the World Premier International Research Center Initiative (WPI), MEXT, Japan. 

\medskip
\bibliographystyle{angew}
\bibliography{AM_Ref}

\end{document}